\documentclass{article}
\usepackage{mrs2005,epsfig}
\begin{document} 
\title{MULTI-WAVELENGTH STUDY OF GALAXY ROTATION CURVES AND ITS APPLICATION TO COSMOLOGY}

\author{Am\'{e}lie Saintonge}
\affil{Cornell University}
\author{Christian Marinoni}
\affil{Centre de Physique Th\'{e}orique, Universit\'{e} de Provence}
\author{Karen L. Masters, Riccardo Giovanelli, Martha P. Haynes}
\affil{Cornell University}
\author{Thierry Contini}
\affil{Observatoire de Toulouse}

\begin{abstract} 
Rotation information for spiral galaxies can be obtained through the observation of different 
spectral lines. While the H$\alpha$($\lambda 6563 \AA$) line is often used for galaxies with low to 
moderate redshifts, it is redshifted into the near-infrared at $z>0.4$. This is why most high redshift surveys rely on the [OII]($\lambda 3727 \AA$)
 line. Using a sample of 32 spiral galaxies at $0.155 < z < 0.25$ observed simultaneously in 
both H$\alpha$ and [OII] with the Hale 200 inch telescope, the relation between velocity 
widths extracted from these two spectral lines is investigated, and we conclude that $H\alpha$ derived velocities can be reliably compared to high $z$ [OII] measurements.  The sample of galaxies is then used along with VIMOS-VLT 
Deep Survey observations to perform the angular diameter - redshift test to find constraints 
on cosmological parameters ($w$, $\Omega_{M}$, $\Omega_{\Lambda}$).  The test makes it possible to discriminate between various cosmological models, given the upper limit of disc size evolution at the maximum redshift of the data set, no matter what the evolutionary scenario is.
\end{abstract} 
 
\section{Introduction} 
With the new generation of telescopes and instruments, such as multi-object spectrographs and large field of view camera, it now becomes possible to perform large, deep redshift surveys.  Using these data, the structure and geometry of the Universe can be studied through new channels, namely geometrical tests such as the angular diameter test (angular size - redshift relation).  To perform such geometrical tests, a population of objects that can be followed through redshift space needs to be identified.  The angular diameter test requires a standard rod to be defined, the angular size of which is then to be measured at various epochs.  The objects taken to serve that purpose could be as diverse as galaxies, clusters of galaxies, or dark matter halos.  In this study, the relation between the physical size of a disk and its rotation velocity is used \cite{tf77}.  By virtue of this relation, selecting a population of objects with a given rotation velocity is equivalent to selecting them by their physical size \cite{marinoni}.  Spectroscopically selecting the photometric standard rods also has the advantage of minimizing selection effects due to the Malmquist bias.

Most ground-based studies of galaxies at high redshift rely on the [OII]$\lambda 3727$ \AA\ line, including the VIMOS-VLT Deep Survey (VVDS)\cite{lefevre} and the DEEP2 Redshift Survey \cite{davis01}, since H$\alpha$ used locally is redshifted into the infrared at about $z>0.4$.  Specifically, the VVDS redshift survey will measure [OII] line widths up to $z \sim 1.4$.  In order to compare data sets of local and distant galaxies, it is therefore necessary to understand how rotation velocities extracted from different lines relate, in this case H$\alpha$ and [OII].  The results are then used to calibrate the angular diameter test that is performed on data from the VVDS.

\section{Optical Spectroscopy of Spirals at $0.155<z<0.25$} 
Using the Hale 200 inch telescope during the course of three observing runs between 2003 March and 2004 February (total of 12 nights, about 50\% of which were lost to bad weather), long-slit spectra were obtained for a sample of 32 spiral galaxies.  The galaxies were chosen in the area covered by the Data Release 1 of the {\it Sloan Digital Sky Survey} \cite{stoughton}, and selected by their [OII]$\lambda 3727$ equivalent width and apparent magnitude.  The galaxies were also required to be in the redshift range $0.155<z<0.25$, so that using the Double Spectrograph \cite{okegunn}, both H$\alpha \lambda$6563\AA\ and [OII]$\lambda 3727$\AA\ (thereafter, H$\alpha$ and [OII]) emission lines could be observed simultaneously.  

The spectral resolution on the red side of the spectrograph is 0.65 \AA\ pixel$^{-1}$ with 24$\mu$m pixels and the spatial scale is of 0\farcs468 pixel$^{-1}$. The CCD camera had a size of 1024$\times$1024 pixels, and a 1\arcsec$\times$128\arcsec\ slit was used (except for a few galaxies where the 2\arcsec$\times$128\arcsec\ slit was used due to very poor seeing conditions).  The blue camera  used a 2788$\times$512 chip (15$\mu$m pixels), giving a spatial resolution of 0\farcs390 pixel$^{-1}$, as well as a 0.55 \AA\ pixel$^{-1}$ spectral resolution.  The exposure time was of 3600 s for most galaxies, though it was extended to 4800 s for a few sources when the flux in the [OII] line was very low.

\section{Rotation Velocities}
The rotation curves were modeled in two steps.  First, a Universal Rotation Curve\cite{urc} was fitted and used to determine the velocity and radial offsets required to get the best folding of the rotation curves.  Secondly, a Polyex model \cite{polyex} was applied to the folded data.
Using these fits, the rotational velocity of each galaxy was determined, both for its H$\alpha$ and [OII] lines.  The velocity adopted for a given rotation curve is the value of the Polyex fit at a radius corresponding to $r_{83}$, the radius containing 83\% of the light of the galaxy.  There is good agreement between the two sets of measurements, especially for $V_{rot}<$220 km s$^{-1}$ (for more details and figures, see Saintonge et al. 2006 \cite{paperI}).  This is a first clue that the [OII] derived velocities can be reliably compared to H$\alpha$ velocities.  

Since rotation curves cannot be traced out for the high redshift galaxies of the VVDS sample, a method based on velocity histograms is used.  We apply the technique to the low redshift sample for which rotation curves are also available to establish reliability.  The velocity histograms are built by collapsing the two-dimensional images along the spatial direction in order to form one-dimensional spectra.  A Gaussian was fitted to each of the H$\alpha$ emission lines.  The full width at half maximum (FWHM) of that Gaussian was converted into the rotation velocity of the galaxy.  For the [OII] line, which is in fact a doublet, a two Gaussian model is fitted.  There is a good correlation between the techniques, such that the velocities derived from the H$\alpha$ rotation curves, which are the most accurate for our low redshift data, can be used to select objects for the angular diameter test in a way that is consistent with the selection of objects from the VVDS survey.

\section{The Angular Diameter Test}
The angular diameter test can discriminate between cosmological models by tracing the apparent angular diameter of galaxies through redshift space.  In this case, the velocity-diameter relation is used to select standard rods, for which the effect of galaxy evolution needs to be untangled from that of geometric evolution.  It is possible to infer cosmological information knowing a priori only the upper limit value for disc evolution at the maximum redshift of the data set,$z_{max}$, no matter what the specific evolutionary scenario is (see Marinoni et al. 2006 for details).  One can even construct a self-consistent cosmology evolution plane where to any given chosen evolutionary upper limit in diameters at $z_{max}$ corresponds a specific region of the cosmological parameter space.  Vice versa, given a cosmology, one may directly determine the evolution in magnitude and size of the selected sample of galaxies.  

Therefore, this diagram may be used to directly detect, in a fully geometric way, the presence of dark energy in the universe, in a way that is complementary to the use of Supernovae.  In particular, it is found that if discs were less than 30\% smaller at $z=1.5$ than at present epoch, then an Einstein-de Sitter critical universe ($\Omega_m=1$) may be geometrically discriminated from a $\Lambda$CDM cosmology.

\vfill 

\begin{thebibliography}{}{

\bibitem{tf77} Tully, R.B., \& Fisher, J.R. 1977, A\&A, 54, 661
\bibitem{marinoni} Marinoni,C. \& Le Fevre, O. 2004, Ap\&SS, 290, 195
\bibitem{lefevre} Le Fevre, O. et al. 2005, A\&A, 439, 845
\bibitem{davis01} Davis, M., Newman, J., Faber, S., \& Phillips, A. 2001, in Deep Fields, Proceedings of the ESO/ECF/STScI Workshop, ed. S. Christiani, A. Renzini, \& R.E. Williams (Garching: Springer), 241
\bibitem{stoughton} Stoughton, C., et al. 2002, AJ, 123, 485
\bibitem{okegunn} Oke, J.B., \& Gunn, J.E. 1982, PASP, 94, 586
\bibitem{urc} Persic, M., Salucci, P., \& Stel, F. 1996, \mnras, 281, 27
\bibitem{polyex} Giovanelli, R., \& Haynes, M.P. 2002, \apj, 571, L107
\bibitem{paperI} Saintonge, A., Marinoni, C., Masters, K.L., Giovanelli, R., \& Haynes, M.P., {\it in preparation}
\bibitem{paperII} Marinoni, C., Saintonge, A., Contini, T., Giovanelli, R., \& Haynes, M.P., {\it in preparation}
} 
\end{thebibliography}
\end{document}